\documentclass[a4paper]{jpconf}
\usepackage{graphicx}
\usepackage{amsmath}
\begin{document}
\title{Spin-wave theory for dimerized ferromagnetic chains}

\author{Alexander Herzog$^{1}$, Peter Horsch$^1$, Andrzej M. Ole\'s$^{1,2}$, Jesko Sirker$^1$}

\address{$^1$ Max-Planck-Institut f\"ur Festk\"orperforschung,
              Heisenbergstr. 1, D-70569 Stuttgart, Germany \\
         $^2$ Marian Smoluchowski Institute of Physics,
              Jagellonian University, Reymonta 4, \\ PL-30059 Krak\'ow, Poland}

\ead{a.herzog@fkf.mpg.de}

\begin{abstract}
We describe a Peierls dimerization which occurs in ferromagnetic
spin chains at finite temperature, within the modified spin-wave
theory. Usual spin-wave theory is modified by introducing a
Lagrange multiplier which enforces a nonmagnetic state at finite
temperature. It is shown that this method gives results in
excellent agreement with the density--matrix renormalization group applied to transfer matrices
for dimerized ferromagnetic chains. We study bond correlation
functions $\langle {\bf S}_i\cdot{\bf S}_{i+1}\rangle$ and explore the characteristic features of dimerization in the specific heat.
\end{abstract}

\section{Introduction}
It is a well established fact that a coupling between electronic
and lattice degrees of freedom can drive structural instabilities
of electronic systems. This effect is induced by a gain in
electronic energy surpassing the cost in elastic energy due to the
lattice distortion. For instance, in the Peierls instability, a
static lattice distortion turns a one-dimensional (1D) free
electron system into a band insulator \cite{Peierls}. A similar
effect has been observed for antiferromagnetic (AFM) spin chains
coupled to phonons where a gain in magnetic energy drives the
so-called spin-Peierls transition \cite{OrgCuGeO}. A further
example for a Peierls instability, where a ferromagnetic (FM)
spin-chain shows a periodic modulation (dimerization) of the
magnetic exchange in a certain finite temperature region was
observed in YVO$_3$ \cite{Ulr03} and has recently been discussed theoretically
\cite{PRL91,PRL101}. This effect is induced by a coupling of the
spin and orbital degrees of freedom \cite{Jesko}.

In this paper we highlight the differences of the spin-Peierls
transition in AFM and FM spin chains and discuss the dimerized FM
spin chain within the framework of modified spin-wave theory
(MSWT). We show that this method yields results in excellent
agreement with numerical data \cite{PRL101}. In particular, we
consider the
specific heat and bond correlation functions. In addition
we give an analytical expression for the free energy at low
temperatures and small dimerization allowing for a quantitative
understanding of the occurrence of dimerization in FM spin chains.

\section{Spin-Peierls effect in spin chains}
First, we investigate under which circumstances a Peierls
dimerization may occur in FM spin chains. It will be also
instructive to discuss briefly the Peierls dimerization in AFM
spin chains to highlight the differences between the FM and AFM
case.

Our starting point is a coupling of the spin chain to lattice
degrees of freedom. If the Peierls gap exceeds the phonon
frequency it is permissible to neglect the phonon fluctuations.
Within this approximation the Hamiltonian reads
\begin{equation}
\label{H} H=J\sum_{j=1}^N\left\{1+(-1)^j\delta\right\}
\mathbf{S}_{j}\cdot\mathbf{S}_{j+1}+\frac{NK}{2}\delta^2\,,
\end{equation}
where we have defined $\delta\equiv2gu/(Ja_0)$ and $K\equiv\tilde
KJ^2a_0^2/(4g^2)$. Here $J$ is the exchange constant,
$\mathbf{S}_j$ a spin $S$ operator at site $j$, $N$ the number of
sites, and $K$ the effective elastic constant with
$E_\textnormal{el}=N\tilde K u²/2$. The dimensionless parameter
$\delta$ is given by the exchange constant $J$, the spin-phonon coupling constant $g$, the
atomic displacement $u$, and the lattice constant $a_0$. We note
that in spite of the fact that the model \eref{H} is 1D, the
static, mean-field treatment of the three-dimensional phonons
allows for a finite temperature phase transition if $\delta(T)$ is
treated as a thermodynamical degree of freedom.
\begin{figure}[t]
    \includegraphics[scale=1.2]{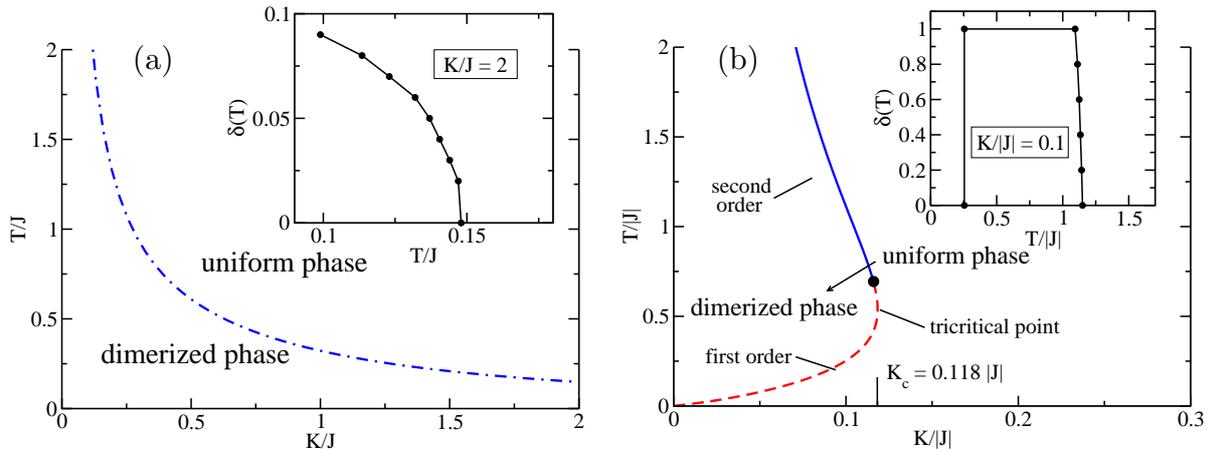}
\caption{Phase diagrams obtained from equation (1) using TMRG for:
(a) the dimerized AFM $S=1/2$ Heisenberg chain --- the inset shows
$\delta(T)$ with $K/J=2$, (b) the dimerized FM $S=1/2$ Heisenberg
chain --- the inset shows $\delta(T)$ with $K/|J|=0.1$.}
\label{PFM}
\end{figure}

For the 1D Heisenberg antiferromagnet it was shown that
$E_\textnormal{mag}\sim-\delta^{4/3}$ \cite{CrFi}. Hence the cost
in elastic energy due to a distortion of the lattice is outweighed
by the contribution stemming from the magnetic part of the
Hamiltonian. This leads to a spin-Peierls transition for every
value of $K$ at sufficiently low temperature. The phase diagram
obtained within the density-matrix renormalization group applied
to transfer matrices (TMRG) \cite{TMRG} is shown in \fref{PFM}(a)
for the dimerized AFM $S=1/2$ Heisenberg chain \eref{H}. The phase
transition is of second order. The inset shows the parameter
$\delta$ as a function of temperature.

In \fref{PFM}(b) the phase diagram for the dimerized FM chain
obtained by TMRG is shown. We observe that a dimerized phase
exists only at $T>0$ and only if $K/|J|<K_c/|J|\simeq0.118$. We
also report a tricritical point (TCP) at
$(T_\textnormal{TCP}/|J|,K_\textnormal{TCP}/|J|)\simeq(0.696,0.116)$.
For $T<T_\textnormal{TCP}$ and $T>T_\textnormal{TCP}$ the
transition will be first and second order, respectively. This is
shown in the inset of \fref{PFM}(b) where we can see that at the
upper phase boundary $\delta(T)$ evolves continuously.
%

\section{Modified spin-wave theory for dimerized chains}

\begin{figure}
 \includegraphics[scale=1.35]{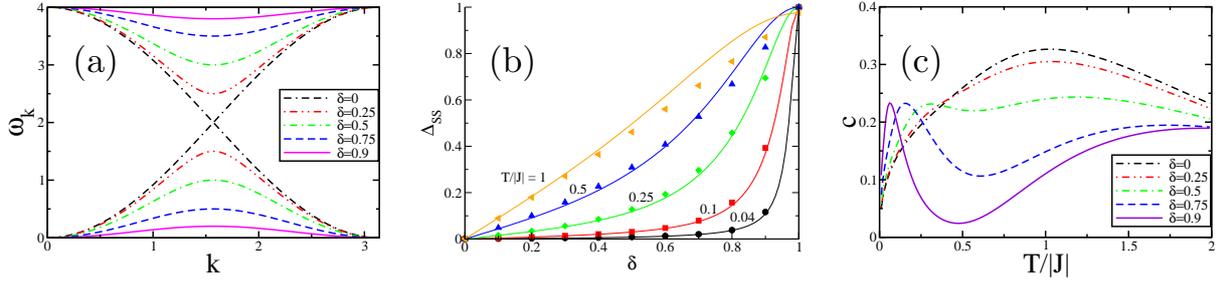}
\caption{MSWT results for $S=1$: (a) Magnon dispersions
$\omega_k^-$ (lower branches) and $\omega_k^+$(upper branches) for
various values of $\delta\in [0,0.9]$. (b) $\Delta_{SS}$ as a
function of $\delta$ for various $T$. The lines (symbols) denote
the MSWT (TMRG) results. (c) Specific heat per site as a function
of temperature for various $\delta$. }\label{sc}
\end{figure}

For the case of a FM chain, we apply a Holstein-Primakoff
transformation onto the magnetic part of Hamiltonian (\ref{H}).
Since the unit cell is doubled for the dimerized chain we have to
distinguish between magnons on two sublattices. Hence we write
\begin{equation}
 S_{j\alpha}^+=\sqrt{2S-a_{j\alpha}^\dagger a_{j\alpha}^{}}\,a_{j\alpha}^{\ }\,,
 \quad S_{j\alpha}^-=a_{j\alpha}^\dagger\,\sqrt{2S-a_{j\alpha}^\dagger a_{j\alpha}^{}}\,,
 \quad S_{j,\alpha}^z=S-a_{j,\alpha}^\dagger a^{\ }_{j,\alpha}\,,
\label{HolsteinPrimakoff}
\end{equation}
where $a^{\ }_{j\alpha}$ and $a_{j\alpha}^\dagger$ are bosonic
annihilation and creation operators at site $j$. The index
$\alpha\in\{e,o\}$ refers to the sublattice for $j$ being an even
($e)$ or an odd ($o$) index, respectively. Retaining only the
lowest order terms we obtain a bilinear Hamiltonian which can be
diagonalized by means of a Bogoliubov transformation.
The diagonalized Hamiltonian reads
\begin{equation}\label{HSWT}
    H_\textnormal{mag}=H_0+\sum_k(\omega_k^-\alpha_k^\dagger\alpha_k^{\ }
    +\omega_k^+\beta_k^\dagger\beta_k^{\ })\,,
\end{equation}
with $H_0=-|J|S^2N$ and the magnon branches
$\omega_k^\pm=2|J|S(1\pm\sqrt{\cos^2k+\delta^2\sin^2k})$ which are
shown in \fref{sc}(a).

The idea of Takahashi's MSWT \cite{Taka} is to retain Hamiltonian
\eref{HSWT} while enforcing the Mermin-Wagner theorem. This is
achieved by minimizing the free energy subject to the constraint
that the magnetization at finite temperature vanishes. To this end
a Lagrange multiplier is introduced serving as a chemical
potential. Applying MSWT for the dimerized FM chain we obtain the
constraint
\begin{equation}\label{constraint}
 S=\frac{1}{N}\sum_k\left\{n_\textnormal{B}(\omega_k^-)
                          +n_\textnormal{B}(\omega_k^+)\right\}\,,
\end{equation}
with
$n_\textnormal{B}(\omega_k^\pm)=\{\exp[(\omega_k^\pm-\mu)/T]-1\}^{-1}$
being the Bose factors. The free energy per site in this approximation reads
\begin{equation}\label{fe}
    f=-|J|S+\mu S-\frac{T}{N}\sum_k\left\{\ln\left[
    1+n_\textnormal{B}(\omega_k^-)\right]
    +\ln\left[1+n_\textnormal{B}(\omega_k^+)\right]\right\}\,.
\end{equation}
For the uniform case ($\delta=0$) it reduces to the free energy
obtained by Takahashi for the 1D Heisenberg ferromagnet
\cite{Taka}.

The free energy can be obtained analytically within the MSWT
framework in the limit $t_\delta\equiv T/\{|J|S(1-\delta^2)\}\ll
1$. To this end we expand equation \eref{constraint}
\cite{Robinson} to obtain
\begin{equation}\label{vau}
 v=t_\delta\left(1+\alpha\sqrt{t_\delta}+\frac{3}{4}\alpha^2t_\delta+\dots\right)\,,
\end{equation}
with $v\equiv-4\mu S^2/T$, and $\alpha\equiv\zeta(\frac{1}{2})/S\sqrt{\pi}$.
Inserting this result into equation \eref{fe} we obtain
$f=-|J|S-T\{\zeta(\frac{3}{2})t_\delta^{1/2}/(2\sqrt{\pi})+\mathcal{O}(t_\delta)\}$.
Hence the gain in magnetic energy is proportional $\delta^2$ as is
the elastic energy, explaining the existence of a critical value
$K_c/|J|$.

Due to the alternating interaction between the spins,
nearest-neighbor correlation functions (NNCF) will also exhibit
alternation in strength. Hence
$\Delta_{SS}\equiv\left<\mathbf{S}_{2j}\cdot\mathbf{S}_{2j+1}\right>
-\left<\mathbf{S}_{2j}\cdot\mathbf{S}_{2j-1}\right>$ serves as an
order parameter for the dimerized chain. In calculating the
corresponding NNCF within MSWT it is essential to also retain
quartic bosonic terms which leads to
\begin{equation}\label{SdotS}
    \left<\mathbf{S}_{2j}\cdot\mathbf{S}_{2j\pm1}\right>=
    \left(\frac{1}{N}\sum_k\left[n_\textnormal{B}(\omega_k^-)
    -n_\textnormal{B}(\omega_k^+)\right]f^\pm_k(\delta)\right)^2\,,
\end{equation}
with
$f_k^\pm(\delta)=(\cos^2k\pm\delta\sin^2k)/(\sqrt{\cos^2k+\delta^2\sin^2k})$.
In \fref{sc}(b) results from MSWT for $\Delta_{SS}$ as a function
of $\delta$ at various temperatures are compared to those obtained
within TMRG. The agreement is good up to $T/|J|\sim1$ and even the
right limit is captured by MSWT for the fully dimerized case.
However, corrections $\delta=1-\epsilon$ are predicted to be of
order $\epsilon^2$ for $\epsilon\ll1$ whereas perturbation theory
exhibits corrections of order $\epsilon$.

With equation \eref{SdotS} it is easy to determine the internal
energy per site reading
\begin{equation}
    u=\frac{1}{2}\,\left\{(1-\delta)\left<\mathbf{S}_{2j-1}\cdot\mathbf{S}_{2j}\right>
    +(1+\delta)\left<\mathbf{S}_{2j}\cdot\mathbf{S}_{2j+1}\right>\right\}.
\end{equation}
Using this result the specific heat per site reads $c=c^++c^-$
with
\begin{equation}
c^\pm=\frac{(1\pm\delta)}{T^2N^2}\sum_{\stackrel{k,k',}{\sigma\in\{\pm\}}}
 (-\sigma)n_\textnormal{B}(\omega_{k'}^\sigma)[1+n_\textnormal{B}
 (\omega_{k'}^\sigma)]\left(\omega_{k'}^\sigma-\mu
 +T\frac{\partial\mu}{\partial T}\right)f_{k'}^\pm\left\{n_\textnormal{B}
 (\omega_k^-)-n_\textnormal{B}(\omega_k^+)\right\}f_k^\pm.
\end{equation}
In \fref{sc}(c) we show the specific heat per site as a function
of temperature for various dimerizations. For $\delta=0$ we find
one broad hump which still is retained if $\delta\ll 1$. However,
at a higher dimerization we observe that the two magnon branches
give distinct contributions leading to a sharp low temperature
peak and a hump at higher temperatures stemming from well
separated magnon excitations, see \fref{sc}(a). The excitations of
the lower magnon band $\omega_k^-$ may occur at low $T$ whereas
those of the higher band $\omega_k^+$ contribute only when $T$ is
high.

\section{Conclusions}
We have discussed  the differences in the occurrence of a
spin-Peierls transition in AFM and FM spin chains. For the FM
chain we have applied MSWT and found that the magnetic energy gain
$\sim-T^{3/2}\delta^2$ directly competes with the cost in elastic
energy. The dimerization order parameter obtained from MSWT
is found to be in excellent agreement with TMRG data.
Finally, we have shown that dimerization leads to a qualitatively
different behaviour of the specific heat.


\ack A M Ole\'s acknowledges support by the Foundation for Polish
Science (FNP) and by the Polish Ministry of Science and Higher
Education under Project No.~N202 068 32/1481.

\end{document}